\documentclass[graybox]{svmult}

\usepackage{mathptmx}       
\usepackage{helvet}         
\usepackage{courier}        
\usepackage{type1cm}        
\usepackage{makeidx}         
\usepackage{graphicx}        
\usepackage{multicol}        
\usepackage[bottom]{footmisc}
\usepackage{amsmath,txfonts
}

\makeindex             
                       
\newcommand*{\okr}{{\stackrel{{\scriptscriptstyle{\mathsf{def}}}}{=}}}
\renewcommand*{\Re}{\mathfrak{Re\,}}
\renewcommand*{\Im}{\mathfrak{Im\,}}
\newcommand*{\lin}{{\rm lin}}  
\newcommand*{\clolin}{{\rm clolin}}

\def\is#1#2{\langle#1,#2\rangle}
\def\Le{\leqslant}
\def\liczp#1{{${#1}^{{\rm o}}$}}

\def\sbar#1{\,\overline{\!#1}}
\def\ulamek#1#2{\mbox{\normalfont$\frac{#1}{#2}$}}

\def\zb#1#2{\{{#1}\colon\ {#2}\}}
\def\nic{\varnothing}

\newcommand*\ddc{\mathcal D}
\newcommand*\hhc{\mathcal H}

\newcommand*\kkc{\mathcal K}
\newcommand*\llc{\mathcal L}

\newcommand*\ssc{\mathcal S}
\newcommand*\xxc{\mathcal X}

\newcommand*\ccb{\mathbb C}
\newcommand*\ddb{\mathbb D}

\newcommand*\rrb{\mathbb R}
\newcommand*\ttb{\mathbb T}

\usepackage{color,colortbl}
\definecolor{mygray}{gray}{.82}

\begin{document}

\title*{Coherence, squeezing and entanglement \\-- an example of peaceful coexistence}
\author{Katarzyna G\'{o}rska, Andrzej Horzela and Franciszek Hugon Szafraniec}
\institute{Katarzyna G\'{o}rska \at H. Niewodnicza\'{n}ski Institute of Nuclear Physics, Polish Academy of Sciences, Division of Theoretical Physics, ul. Eliasza-Radzikowskiego 152, PL 31-342 Krak\'{o}w, Poland, \\ \email{katarzyna.gorska@ifj.edu.pl}
\and Andrzej Horzela \at H. Niewodnicza\'{n}ski Institute of Nuclear Physics, Polish Academy of Sciences, Division of Theoretical Physics, ul. Eliasza-Radzikowskiego 152, PL 31-342 Krak\'{o}w, Poland, \\ \email{andrzej.horzela@ifj.edu.pl}
\and Franciszek Hugon Szafraniec \at Instytut Matematyki, Uniwersytet Jagiello\'{n}ski, ul. \L ojasiewicza 6, 30 348 Krak\'ow, Poland, \\ \email{umszafra@cyf-kr.edu.pl}}
%
%
\maketitle

\abstract{
After exhaustive inspection of bosonic coherent states appearing in physical literature two of us, Horzela and Szafraniec, came in 2012 to the reasonably general definition which relies exclusively on reproducing kernels. The basic feature of coherent states, which is the resolution of the identity,  is preserved though it now achieves advantageous form of the Segal-Bargmann transform.
\\
\noindent
It turns out that the aforesaid definition is not only extremely economical but also puts under a common umbrella  typical coherent states as well as those which are squeezed and entangled. We examine the case here on the groundwork of holomorphic Hermite polynomials in one and two variables. An interesting side of this story is how some limit procedure allows disentangling.}

\section{Coherent states - a smooth introduction}
\label{GHSz:sec:f1}
{\em Coherent states} (CSs in short) constitute a vivid topic in Quantum Optics besides being of interest from the mathematical point of view. This Section provides a short though solid introduction culminating in the fairly recent extension of the notion which has a novel and  pretty interesting application {(cf. Section \ref{GHSz:sec:3})}.
\noindent{As a kind of shorthand to this Section the presentation \cite{GHSz:luminy-fhsz} may serve.}

\subsection{Standard coherent states}\label{GHSz:ssec:f1.1}
{What are coherent states?}  
The    {standard} harmonic oscillator coherent states\footnote{They bear different names like canonical, classical, orthodox, Glauber-Klauder-Sudarshan (GKS in short, \cite{GHSz:glauber, GHSz:klauder1,GHSz:Sudarshan}), etc.  though the most explicative way would be to call them,  as it becomes clearer later, {\em Gaussian} coherent states upon the Gaussian kernel involved in \eqref{GHSz:f2.18.07}.}, {originated in \cite{GHSz:szre},}  are
 simply
\begin{equation}\label{GHSz:f2.18.07} c_{z}\okr
{  {\exp(-|z|^{2}/2)}}\sum_{n=0}^{\infty}\frac{z^{n}}{\sqrt{n!}}h_{n},\quad z\in\ccb,
\end{equation}
with $h_{n}$'s being the Hermite functions 
\begin{equation*}
h_n(x)=2^{-n/2}(n!)^{-1/2}\pi^{-1/4}\E^{-x^2/2}H_n(x),\;\text{ $H_{n}(x)$ is the $n$-th Hermite polynomial.}
\end{equation*}
As the Hermite functions $h_{n}$ are residing in $\llc^{2}{(\rrb)}$ the safest way to consider convergence in \eqref{GHSz:f2.18.07} is  to require it in this space.

 Immediately from the definition \eqref{GHSz:f2.18.07} one usually derives that  such introduced $c_{z}$ are:
 \begin{enumerate}
\item[(a)] normalized;  \label{aaa}
\item[(b)] continuous functions in $z$;  \label{GHSz:con}
\item[(c)] never orthogonal, even more $\is{c_{z}}{c_{w}}=\E^{-{|z-w|}^{2}}$;  
\item[(d)] temporally stable \cite[p.32]{GHSz:jpg}
$$
\E^{\I Ht}c_{z}=\E^{-\I\frac {\omega  t}{2}}c_{\E^{\I \omega t}z},\quad\text{$H$ is the harmonic oscillator Hamiltonian;}
$$
\end{enumerate}
and, last but not least, satisfy the celebrated relation
\footnote{In this section standard Hilbert space notation is employed; there are two exceptions when Dirac's notation is in use: here and on p. \pageref{GHSz:dirac}.}
\begin{enumerate}
\item
[(e)] $I=\int\nolimits_{\ccb}|z \rangle\langle z|\frac{\D^{   {\,2}}z}{\pi}$, $\;$here $|z\rangle$ states for $c_{z}$\label{GHSz:res}.
\end{enumerate}
   \begin{remark}\label{GHSz:r:f1.20.07}{(e)} is customarily called {\em the resolution of the identity}, sometimes referred to as (over)completeness.   
   \end{remark}

In the literature\,\footnote{\;The basic monographs \cite{GHSz:ali,GHSz:jpg,GHSz:klauder-s1,GHSz:klauder-s2,GHSz:perelomov}
of the subject can be completed with other articles like \cite{GHSz:dodonov}, \cite{GHSz:sivakumar} and \cite{GHSz:Zhang}.} one finds three ways of constructing  CSs:   
\begin{enumerate}
\item[(A)]\label{AAA}
as the (normalized) eigenvectors of the annihilation operator\,\footnote{They are in fact in the domain of the closure of annihilation rather than in its domain as usually seems to be thought of; see \cite{GHSz:znojl} for more on the operators.}; \label{GHSz:fA}   
\item[(B)] as the orbit of the vacuum under a square integrable representation of a unitary group;    
\item[(C)] as states which minimize the Heisenberg uncertainty relation.
\end{enumerate}   
It turns out that for the GKS CSs these three lead to the same provided in (B) the group is that of the {\em displacement} operator.

\paragraph{\sc A closer look at the properties {\bf (a)-(e)}}\label{GHSz:ss:f2.20.07}
Property (a) is superfluous, the normalization it serves for can be achieved any time it is needed because CSs are vectors in a Hilbert space and as such they have ``finite" norms. In general normalization may destroy holomorphicity of CSs in the case it may be present ({\em vide} the Segal-Bargmann space).

\noindent Properties (b) and (d) depends on circumstances or in other words on structure of the set which CSs are parametrised by; removing normalizability as suggested above makes  the GKS CSs even holomorphic.

\noindent The angle between CSs calculated in most of the cases supports property (c).

\noindent The resolution of the identity property (e) is our main concern in this Section and different its aspects will be discussed.

\subsection{After 1963}\label{GHSz:ss:f3.1.20.07} Since their rediscovery in 1963 CSs have begun spreading out and a plethora of their different versions rooted in various branches of physics have appeared (for a fairly account of most of  those diversities see e.g. \cite{GHSz:dodonov}, \cite{GHSz:jpg} and \cite{GHSz:klauder-s1}). In particular, {MP}s (this an apparent abbreviation for either Mathematical Physics itself or its admirers), still  keeping in mind the postulates {(a) -- (e)} and following any of the directives {(A), (B) and  (C)}, have been trying to find either generalizations of CSs or to provide any evidence of existence of CSs in various fields of physics.   

Being joined to these efforts we adopt as our starting point to give a precise meaning to what is in Remark \ref{GHSz:r:f1.20.07} and by the way to respond to the call formulated by J. R. Klauder as the Postulate 3 \textit{Completeness and resolution}\footnote{\textit{"A resolution of unity in the Hilbert space $\hhc$ exists as an integral over projection operators onto individual vectors in the (coherent states) set ${\ssc}$."}} in his seminal paper \cite{GHSz:klauder1}. Klauder's approach, later on pushed forward in \cite{GHSz:klauderMPL}\footnote{\textit{"Traditionally, coherent states rely heavily for their construction and analysis on properties of suitable Lie algebra generators appropriate to some specific group. Hence, most of the properties of the coherent states are inherited from the group itself....We entirely set aside any group .... and proceed more generally. We are led to an extremely wide class of coherent states that includes group-defined coherent states as a \underline{small subset}"}(underlined by the authors of the present paper).}, has found its further development in \cite{GHSz:jpg-klauder} where J.-P. Gazeau and J. R. Klauder have proposed to make the following replacements in \eqref{GHSz:f2.18.07}  (we refer to those as to KG CSs {though pretty often they are referred to as {\em nonlinear CSs}} \cite{GHSz:sivakumar}):
\begin{enumerate}
\item
[$ \cdot$]
${h_{n}}'s$ $\mapsto$ arbitrary orthonormal basic vectors in \underbar{some} Hilbert space (in which the would-be CSs have to reside);   
\item
[$\cdot$]
$n!$ $\mapsto$ $\rho_n=\epsilon_{1}\ldots \epsilon_{n} > 0$  in the way which ensures convergence; the sequence $(\epsilon_{n})_{n}$ is usually assumed to be related to the spectrum of the Hamiltonian (describing the physical system under consideration) in a way which guarantees temporal stability and so-called "action identity"; 
\item
[$\cdot$]
$\exp(-|z|^{2}/2)$ $\mapsto$ a suitable normalization factor if any;
\item
[$\cdot$]
${\D^{   {\,2}}z}$ in (e) on p. \pageref{GHSz:con} $\mapsto$ a rotationally invariant measure on $\ccb$ with a radial factor coming from (and solving) the Stieltjes moment problem\footnote{ This takes place for a vast majority of examples present in the literature (see e.g. \cite{GHSz:Klaudermom, GHSz:penson1, GHSz:penson2} ).}; one checks that this secures the property (e).
\end{enumerate}

The final step in such realized generalization of the CSs concept is that proposed by J.-P. Gazeau in the Ch.5 of \cite{GHSz:jpg}: CSs are introduced as continuous in $x$ and normalizable linear combinations  
$$c_x\okr\sum_{n\in N}\phi_{n}(x)e_n, \quad x\in X, $$     
where $(e_n)_{n}$ are normalized eigenvectors of a self-adjoint operator $A$ and $(\phi_{n}(x))_n$ is an orthonormal system of functions in $L^2(X, \nu)$ being in one-to-one correspondence with $(e_n)_{n}$  and satisfying $\sum_{n}|\phi_{n}(x)|^2<\infty$ for all $x\in X$ (normalization condition). This allows to get the property (e) 
$$I=\int\nolimits_{X}|x \rangle\langle x|\nu(dx)$$
where, as previously,  $|x\rangle$ states for $c_{x}$.%
\begin{svgraybox}Everything happens in the \underbar{presence} of a \underbar{measure} which makes the resolution of the identity possible; this is out of any discussion there. Even if a measure exists it may not be unique and if the latter happens a plenty of non-rotationally invariant measures \underbar{always} have to appear. More than, no measure may exist though suitably understood resolution of identity can be done which makes the new (generalized) CSs good sense. This may be painful and in this Section we propose a cure for that.
\end{svgraybox}

\subsection{Reproducing kernel Hilbert space - instructional material}\label{GHSz:ss:f1.21.07
}
  
The tool is the {\em reproducing kernel Hilbert space} (RKHS in short) approach a gentle introduction to which follows. 

\medskip

\begin{center}\colorbox{mygray}{A set $X$ given,} call it {\em basic} or {\em supporting}.\end{center}

\medskip

{Given a Hilbert space $\hhc$ of complex functions on $X$ and a function $K: X\times X \mapsto \ccb$;} 
$(\hhc,K)$ is called a {\em RKHS couple} if
\begin{enumerate}
\item
[${\cdot}$]
$K_{x}\okr K(\,\cdot\,,x)\in\hhc,\quad x\in X$;
\item
[${\cdot}$]
$f(x)=\is f{K_{x}},\quad f\in\hhc,\;x\in X.$
\end{enumerate}
The second fact is just referred to as the celebrated {\em reproducing kernel property}. Therefore, we call $K$ the {\em reproducing kernel}. There is a list of properties coming out of this definition and each of them may work for construction the couple, cf. \cite{GHSz:wuj1}\,\footnote{\;This item as well as the other (\cite{GHSz:wuj2}) contains excerpts from \cite{GHSz:wuj}; the proofs are contained in the latter.}. Among them one finds positive definiteness of the kernel $K$ and boundedness of the evaluation functional on $\hhc$.

 Fundamental for us, however, is {Zaremba's formula (\cite{GHSz:zar2})} and its consequences.
Given a sequence $(\Phi_{n})_{n=0}^{\infty}$ of complex functions on $X$ such that
\begin{svgraybox}
{\begin{equation}\label{GHSz:f1.21.07}
{\sum_{n=0}^{\infty}|\Phi_{n}(x)|^{2}<+\infty,\quad x\in X.}
\end{equation}}\end{svgraybox}
  {Then}
$$ K(x,y)\okr\sum_{n=0}^{\infty}\Phi_{n}(x)\overline{\Phi_{n}(y)},\quad x,y\in X,$$
is a  positive definite kernel and, consequently, due to Moore-Aronszajn's construction, see \cite{GHSz:aron} or \cite{GHSz:wuj1} for instance, it uniquely determines its partner, denoted by $\hhc_{K}$ further on, so that they both together constitute a reproducing kernel couple.
This may serve as a very \underbar{practical} way of constructing RKHS.

 \paragraph{\sc What is the role played by the functions $\Phi_{n}$} 

{\liczp 1}. It follows from the Schwarz inequality applied to \eqref{GHSz:f1.21.07} that for any $\xi=(\xi_{n})_{n=0}^{\infty}\in \ell^2$: 

\begin{itemize}
\item{the series
   $$\sum_{n=0}^{\infty}\xi_n \Phi_n(x)$$
   is absolutely convergent for any $x$;} 
   
   \item{the function
   $${f_\xi}:{x}\to{\sum_{n=0}^{\infty}}\xi_n
\Phi_n(x)   
$$
is in $\hhc_{K}$ with
$\|f_\xi\|\le\|\xi\|_{\ell^2}$; }

\item{moreover, 
$\sum_{n}\xi_n \Phi_n$ is convergent in
$\hhc_{K}$ to $f_\xi$. }
\end{itemize}

\noindent In particular
$\sum_{n}\overline{\Phi_n(x)}\Phi_n$ is convergent in  $\hhc_{K}$ to $K_x$, the functions $\Phi_n$ are in $\hhc_{K}$ and $\|\Phi_n\|\le1$.

\medskip\medskip

\noindent{\liczp 2}.\label{GHSz:li2}
  The sequence $(\Phi_n)_{n=0}^{\infty}$ is \underbar{always} \underbar{complete} \,\footnote{\;Notice completeness of $(\Phi_n)_{n}$ appears {\em a posteriori}.}\,\footnote{\;{\em Complete} or {\em total} means the closed linear span $\clolin\zb{\Phi_{n}}{n=0,1,\dots}$ is $\hhc_{K}$. This is equivalent to saying that the only function in $\hhc_{K}$ orthogonal to  all the $\Phi_{n}$'s is $0$.} in $\hhc_{K}$. Moreover, the following facts are equivalent
   \begin{enumerate}
   \item[(i)] $\xi\in\ell^2$ and
$\sum_{n}\xi_n \Phi_n(x)=0$ for every
$x$ yields $\xi=0$;
   \item[(ii)] the sequence $(\Phi_n)_{n}$ is {\em orthonormal} in $\hhc_{K}$.
   \end{enumerate}
   
	\begin{svgraybox}
	It is recommended  to notify that Zaremba's construction guarantees always {\bf completeness} of the sequence  $(\Phi_{n})_{n=0}^{\infty}$ in $\hhc_{K}$; it is an intrinsic feature of the approach. {\bf Orthonormality} of  $(\Phi_{n})_{n=0}^{\infty}$, on the other hand, requires additional effort as \liczp 2 above shows. If the latter happens, $(\Phi_{n})_{n=0}^{\infty}$ must necessarily be a Hilbert space basis of $\hhc_{K}$.
	\end{svgraybox}

\subsection{Horzela-Szafraniec's CSs and the Segal-Bargmann transform}\label{GHSz:ss:f2.21.07}

\paragraph{\sc Horzela-Szafraniec's CSs}
The \underbar{only} data, which the Horzela-Szafraniec procedure \cite{GHSz:H-Sz1,GHSz:H-Sz2} requires besides the supporting set $X$, are
\begin{svgraybox}\begin{itemize}
\item
 a sequence $\boldsymbol{\Phi}\okr(\Phi_{n})_{n=0}^{\infty}$ of  functions on $X$ such that \eqref{GHSz:f1.21.07} holds;
 \item
  a separable Hilbert space $\hhc$ (it can be thought of as a surrogate of the \underbar{state} space).
  \end{itemize}\end{svgraybox}
 Now let $K$ be the  kernel on $X$ got via Zaremba's construction and  $\hhc_{K}$ its RKHS. Fix an   orthonormal basis   $\boldsymbol{e}\okr (e_{n})_{n=0}^{\infty}$ in $\hhc$. Introduce the  family $\{c_{x}\}_{x\in X}$ 
 \begin{equation}\label{GHSz:23/07-2}
c_{x}\okr\sum_{n=0}^{\infty}{\Phi_{n}(x)}e_{n}\quad x\in X.
\end{equation}   
 
  \begin{svgraybox}
 We do not suppose for a while that  $\boldsymbol{\Phi}\okr(\Phi_{n})_{n=0}^{\infty}$ are orthonormal.
 \end{svgraybox}

Let us mention that positive definiteness of $K$, or rather some Schwarz type inequalities which follow, guarantees continuous or holomorphic dependence on $x$ of so introduced $c_{x}$ according to circumstances (cf. \cite{GHSz:wuj1}); this refers to (b) on p. \pageref{GHSz:con}.

\paragraph{\sc The Segal-Bargmann transform}   
The transform 
\begin{equation}\label{GHSz:f4.21.07}
Bh\okr\sum\nolimits_{n=0}^{\infty}\Phi_{n}\is h{e_{n}}_{\hhc},\quad h\in\hhc,
\end{equation}
is well defined and maps
$\hhc \mapsto{\hhc_{K}}$  (notice $Be_{n}=\Phi_{n}$); convergence in \eqref{GHSz:f4.21.07} is that of $\hhc_{K}$. It is a contraction with a dense range.

Due to the reproducing property we have
\begin{equation}\label{GHSz:f6.22.07}
 {(Bh)(x)}=\is{Bh}{K_{x}}_{\hhc_{K}}=  {\sum\nolimits_{n=0}^{\infty}\Phi_{n}(x)\is h{e_{n}}_{\hhc}}=\is {c_{x}}h_{\hhc}, \quad h\in\hhc,\;x\in X,
\end{equation}
 with the convergence being uniform on those subsets of $X$ on which $K(x,x)$ is bounded.
  
Moreover if  $(\Phi_{n})_{n=0}^{\infty}$ is an orthonormal basis then \eqref{GHSz:f4.21.07} and the Parseval formula yields
 \begin{equation}\label{GHSz:f5.21.07}
 \is{Bh}{Bg}_{\hhc_{K}}=\is{h}{g}_{\hhc},\quad g,h\in \hhc;
 \end{equation}
 hence $B$ is   {unitary}.
\begin{theorem}\label{GHSz:ft1.21.07}

 The following three facts are equivalent
 \begin{itemize}
\item
the transform $B$ is   {unitary};   
\item
the family $\{c_{x}\}_{x\in X}$ is   {complete};
\item
the sequence $(\Phi_{n})_{n=0}^{\infty}$ is orthonormal in $\hhc_{K}$.
\end{itemize}   
\end{theorem}

In the GKS prototype, that is when 
$$\Phi_{n}=\frac {z^{n}}{\sqrt{n!}} \text{ or } K(z,w)=\E^{z\sbar w}$$ 
and $e_{n} = h_{n}$ are Hermite functions,  the transform $B$ becomes that of Segal-Bargmann \cite{GHSz:VBargmann61,GHSz:hall}.

\begin{svgraybox}
\begin{minipage}{330pt}
Now it is a right time to declare: call the vectors (states) $c_x$, ${x\in X}$ {\em Horzela - Szafraniec coherent states}
\,\footnote{\;Nicknamed HSz CSs} if  they are given by \eqref{GHSz:23/07-2} and the family  $\{c_x\}_{x\in X}$ is complete in $\hhc$.
\end{minipage}\end{svgraybox}
   
    \paragraph{\sc Horzela-Szafraniec coherent states back and forth\label{GHSz:CS}}
    Universality of our definition of coherent states can be enhanced by the fact which follows
    \begin{svgraybox}
   \begin{proposition}\label{propro} Let  $\hhc$ be a Hilbert space and $(e_n)_{n=0}^{\infty}$ be an orthonormal base in it (one can think of it as the Fock basis). Any family of vectors (states) $\{c_x\}_{x\in X}$ in $\hhc$ becomes a family of coherent states in a sense of Horzela-Szafraniec with respect of the {\bf uniquely determined} reproducing kernel $$K(x,y)=\sum_{n}\is{c_x}{e_n}\overline{\is{c_y}{e_n}}\quad\quad x,y\in X.$$ 
   \end{proposition}
   This implies that  all the coherent states already present in the literature (cases like A, B, C on page \pageref{AAA}) fit in with the Horzela-Szafraniec class; the states mentioned at the end of Section \ref{GHSz:sec.30} are within this class too.
   
    \noindent In particular the Segal-Bargmann transform is valid and Theorem \ref{GHSz:ft1.21.07} holds. 
    \end{svgraybox}  
    Once more, notice that the resolution of the identity (e), p. \pageref{GHSz:res}, (which in our approach, as will be seen explicitly very soon, turns into the Segal-Bargmann transform) is an {\em a posteriori}
fact coming out of the construction, not an {\em a priori} postulate.

    \paragraph{\sc Resolution of the identity for malcontents\label{GHSz:dirac}} 
  \begin{definition} \label{GHSz:def1}  
If $X$ is a (subset of a) topological space and  there is a positive measure $\mu$ on the completion $\overline X$ of $X$ such that $\hhc$ is \underbar{embedded} \underbar{isometrically} in ``a natural way" in $\llc^{2}(X,\mu)$ we say that $(\hhc,K)$ is   {\em integrable}. 
\end{definition}
Let us emphasise that there are non-integrable RKHSs, look at p. \pageref{GHSz:p23/07-1}.    \\

 If $\mu$ is \underbar{any} measure which makes integrability of RKHS possible then\,\footnote{\;Notice Dirac's notation is used  for the second time in this section.}
   \begin{align}
\begin{split}\label{GHSz:111}
\colorbox{mygray}
{$\langle h|\int_{X}|x\rangle\langle x|\,\mu (\D x)| g\rangle $}
&=
\int\nolimits_{X}\langle h| x\rangle\langle x| g\rangle\,\mu(\D x)
\\
& {=}\int\nolimits_{X}\is h {c_{x}}_{\hhc} \overline{\is g{c_{x}}_{\hhc}}\,\mu(\D x)
\\ 
&\stackrel{\eqref{GHSz:f6.22.07}}{=}\int\nolimits_{X}{(Bh)(x)}\overline{(Bg)(x)}\,\mu(\D x)\\
&=\is{Bh}{Bg}_{{\llc}^{2}(X,\mu)} 
\stackrel{{\rm Definition\,} \ref{GHSz:def1}}{=}  \is{Bh}{Bg}_{\hhc_{K}} \\
&\stackrel{\eqref{GHSz:f5.21.07}}{=} 
\is hg_{\hhc}=\colorbox{mygray} 
{$\langle{h}|{g}\rangle$}.
%
\end{split}
\end{align}
\begin{svgraybox}
\begin{minipage}{330pt}
Resolution of the identity, the key feature of CSs, has been rescued\,\footnote{\;The grayish boxes in \eqref{GHSz:111} read together uncover the resolution of the identity as exposed in (e) on p.\pageref{GHSz:res}.} 
in the full glory! Now it bears the name {\bf Segal-Bargmann transform.} \\ 

\noindent
All this justifies once more the use of term {\bf coherent states} for the family $\{c_{x}\}_{x\in X}$. 
\end{minipage}
\end{svgraybox}

  \subsection{The measure -- to be or not to be?}\label{GHSz:sf1.22.07}  
  
Three possibilities for the family  $\{c_{x}\}_{x\in X}$  of  CSs may happen. 
 \paragraph{\sc $\hhc_{K}$ is integrable and the measure is   {unique}.
}
  Here is a list of assorted cases.

   $\blacktriangle$ {Standard  CSs}

 $$\Phi_{n}=\ulamek{z^{n}}{\sqrt{n!}}, \quad z\in\ccb \text{, }\quad e_{n}=\text{Hermite functions and $K(z,w)=\E^{z\sbar w}$}.$$ 
  
$\blacktriangle$ {van  Eijndhoven--Meyers' orthogonality}, cf. p. \pageref{GHSz:sec:2}.
 
 $\blacktriangle$ 
 CSs on the unit circle. They come from the Szeg\H{o} kernel; here
 $$\Phi_{n}(z)=\sqrt{\ulamek {1}{2\pi}}z^{n}, \quad \text{ with } K(z,w)=\ulamek 1{2\pi}(1-z\sbar w)^{-1},\quad |z|,|w|< 1,
 $$
and $\hhc_{K}$ is the space of holomorphic functions on the open unit disk $\ddb$ which is customarily named after Hardy\,\footnote{\;{Notice there a bifurcation of names in this case.}}. The corresponding measure is supported on the unit circle $\ttb\subset \overline \ddb$, cf. the definition of integrability on p. \pageref{GHSz:def1}.

$\blacktriangle$ 
Bergman kernels. Here
$$\Phi_{n}(z)=\sqrt{\ulamek{n+1}{2\pi}}z^{n}, \quad \text{ with } K(z,w)=\ulamek 1{2\pi}(1-z\sbar w)^{-2},\quad |z|,|w|< 1,
 $$
and  the corresponding space $\hhc_{K}$ again {is} composed of holomorphic functions on the open unit disk $\ddb$.

 $\blacktriangle$  $q$-Gaussian CSs for $-1<q<1$; the corresponding $q$ moment problem is determined and the operators appearing in the $q$-oscillator are \underline{bounded}, see \cite{GHSz:fhsz3} and the references therein. 

   \paragraph{\sc $\hhc_{K}$ is integrable and the measure is not unique}
   Two cases for the time being.
    
    $\blacktriangle$
    Typical providers are indeterminate moment problems or rather orthonormal polynomials coming from them. If $(\Phi_{n})_{n}$ is such a sequence of polynomials then the well known consequence is that it satisfies \eqref{GHSz:f1.21.07}. As already shown any of the orthogonality measures appearing in this problem works well for the resolution of the identity to be satisfied. It may create problems for further use of this property. Our construction of CSs and, in particular, of the Segal-Bargmann transform opens a way of overcoming obstacles which may appear.
    
    $\blacktriangle$
    The case $q>1$ is also considered in \cite{GHSz:fhsz3}. In \cite{GHSz:explicit} two different kinds of orthonormal bases and their RKHS's are given explicitly: one measure is absolutely continuous with the Lebesque measure on $\ccb$, and the other is supported on a countable family of circles tending both to zero and infinity. Needless to say, if $q\to 1+$ both RKHS converge do the GKS picture.

\paragraph{\sc $\hhc_{K}$ is   {not} integrable} \label{GHSz:p23/07-1}
 
 $\blacktriangle$ The Sobolev space on $[0,1]$, which is a RKHS, cf. \cite[p. 321]{GHSz:berl}, is recognized as an example of non-integrable RKHS; to see this perform an argument with logarithmic convexity like on p. \pageref{GHSz:flcon}.    
 
 \noindent$\blacktriangle$ 
 Consider now
$$
\Phi_{n}(z)\okr\frac{n!}{z(z+1)\cdots(z+n)}.
$$ Then
\begin{align}
\begin{split}
K(z,w)=&\sum\nolimits_{n=0}^{\infty}\frac{n!}{z(z+1)\cdots(z+n)}
\frac{n!}{\sbar w(\sbar w+1)\cdots(\sbar w+n)}
\\
=&
\,{}_{3\,}\!F_{2}\left(
   1,1,1;z+1,\sbar w+1;1
   \right),\quad \Re z,\Re w>1/2.
   \end{split}\notag
\end{align} 
and the space $\hhc_{K}$ is \underbar{not} integrable over $X=\zb{(z,w)}{\Re z,\Re w>1/2}$ though HSz CSs make sense. This is a kind of surprising, thought-provoking example, see \cite{GHSz:klop}.
 
Notice that $\hhc_{K}=\zb{\sum_{n}\xi_{n}\Phi_{n}}{(\xi_{n})_{n}\in\ell^{2}}$ is the Segal-{Bargmann} type space of holomorphic functions on $\zb{(z,w)}{\Re\, z,\Re\, w>1/2}$. 
 
  \noindent$\blacktriangle$ 
  See the graybox on the p. \pageref{GHSz:man1}.
  
  \paragraph{\sc Another look at KG CSs}
Suppose a sequence $(k_{n})_{n=0}^{\infty}$ of positive numbers (cf. the second item in the list of the KG postulates)  is given such that 
$$X=\zb{z\in\ccb}{\sum_{n}k_{n}^{2}|z|^{2n}<+\infty}\neq\nic.$$   This set is rotationally invariant and so is the kernel
$$
K(z,w)\okr\sum\nolimits_{n}k_{n}z^{n}\sbar w^{n},\quad z,w\in X,
$$
which is well defined due to the Schwarz inequality.
Because $K$ is positive definite, we get RHKS $\hhc_{K}$. 
Furthermore,   
the monomials $\Phi_{n}\okr k_{n}^{1/2}z^{n}$ are  \underbar{orthonormal}\,\footnote{\;This is due to the fact that the sum appearing in \liczp 2, (ii), p. \pageref{GHSz:li2} is holomorphic.} in $\hhc_{K}$. Consequently,
\begin{equation}\label{GHSz:f1.27.07}
\|\Phi_{n}\|_{\hhc_{K}}=1=k_{n}^{1/2}\|z^{n}\|_{\hhc_{K}}.
\end{equation}

Suppose for a while $\hhc_{K}$ is integrable and  using \eqref{GHSz:f1.27.07} write\label{GHSz:flcon}
\begin{align}
\begin{split}
k_{m+n}^{-2}&=\big(\int_{X}|z^{m+n}|^{2}\mu(\D z)\big)^{2}=\big(\int_{X}|z^{2m}||z^{2n}|\mu(\D z)\big)^{2}\\
& \stackrel{{\rm Schwarz}}{\Le}  k_{2m}^{-1}k_{2n}^{-1}\int_{X}|k_{2m}^{\frac 12}z^{2m}|^{2}\mu(\D z)\int_{X}|k_{2n}^{\frac 12}z^{2n}|^{2}\mu(\D z)
=k_{2m}^{-1}k_{2n}^{-1}.
\end{split}\notag
\end{align}

What we have got from the above heuristic reasoning is
$$k_{m+n}^{-2}\Le k_{2m}^{-1}k_{2n}^{-1},$$
which is just \underbar{logarithmic} \underbar{convexity} of $(k_{n}^{-1})_{n}$.
Therefore logarithmic convexity  is a \underbar{necessary} condition  for integrability; it is important to know that.
\begin{svgraybox}
Manipulating $(k_{n}^{-1})_{n}$ to break down logarithmic convexity may provide at once  examples of non-integrable $\hhc_{K}$.
\end{svgraybox}

Start now from a measure $\nu$ representing a Stieltjes moment sequence\label{GHSz:man1}
$$
a_{n}=\int\nolimits_{0}^{+\infty}x^{n}\nu(\D x),\quad n=0,1,\ldots,
$$ 
and define  the rotationally invariant measure  $\mu$ on $\ccb$
   \begin{equation*}
   \mu(\sigma)\okr(2\pi)^{-1}\int_0^{2\pi}\int_0^{+\infty}\chi_\sigma(r\E^{\I
\varphi})\nu(\D r)\D \varphi,\quad \sigma \text{ Borel subset of } \mathbb C,
   \end{equation*}
   where $\chi_{\sigma}$ is the characteristic (indicator function) of $\sigma$. 
   
  If $k_{n}^{-1}\okr \ulamek 1{2\pi}a_{2n}$ then because
   $$
\int\nolimits_{\ccb}|z|^{2n}\mu(\D z)=\int\nolimits_{0}^{+\infty}r^{2n}\nu(\D r), 
$$
$(\Phi_{n})_{n}$, $n=0,1,\dots$, are \underbar{orthonormal} in $\llc^{2}(\ccb,\mu)$ as well. Hence the inclusion is isometric and $\hhc_{K}$ is integrable.

\medskip

\noindent\colorbox{mygray}{Warning:}
if the Stieltjes moment problem for $(a_{n})_{n}$ is indeterminate, besides rotationally invariant $\mu$'s, non-rotationally invariant measures exist too - despite the fact that the kernel itself is rotationally invariant, cf. \cite{GHSz:kro} and \cite{GHSz:fhsz3}. 
This never happens when $\nu$ is determinate, in particular if it has a compact support.

\noindent It is creditable to suggest here $q$-moments: determinate  if $0<q\Le1$ and indeterminate if $q>1$ which covers both cases \cite{GHSz:explicit,GHSz:fhsz3}.

\begin{remark}\label{GHSz:f.r1.27.07}
Introduce the sequence $\sigma_{0}\okr k_{0}^{-1/2}$, $\sigma_{n}\okr \ulamek {k_{n}}{k_{n+1}}$ and define the weighted shift operators, cf. \cite[p.146]{GHSz:jpg}
$$
a_{+}\Phi_{n}\okr \sqrt{\sigma_{n+1}}\Phi_{n},\quad a_{-}\okr\sqrt{\sigma_{n}}\Phi_{n-1},\quad a_{-}\Phi_{0}\okr 0.
$$
They can viewed as generalized when compared with the standard definition of the creation and annihilation operators, p. \pageref{GHSz:fA} is put in an application. This holds independently of whether $\hhc_{K}$ is integrable or not.
\end{remark}

\section{Holomorphic Hermite polynomials}\label{GHSz:sec:2}

The holomorphic Hermite polynomials in one and two variables, as well as  holomorphic Hermite functions determined by them, will be our main tool extensively used in the next Sections to construct coherent states. This Section serves as a kind of technical introduction and revokes the formulae derived and proved in \cite{GHSz:STAli14a} and \cite{GHSz:KGorska17-arXiv}. 

\subsection{Holomorphic Hermite polynomials in a single variable  }\label{GHSz:sec:2}
The Hermite polynomials 
\begin{equation}\label{GHSz:eq2.1}
 H_n(z) = n! \sum_{m=0}^{\lfloor n/2\rfloor} \frac {(-1)^m (2z)^{n-2m}}{m!\; (n-2m)!}, \quad z = x + {\rm i} y,
\end{equation}
are treated here as functions of a single complex variable $z$ and as such they become holomorphic.

\paragraph{\sc van Eijndhoven-Meyers orthogonality}

As is shown in \cite{GHSz:SJLvanEijndhoven90} $H_n(z)$ satisfy the orthogonality relations 
\begin{equation}\label{GHSz:eq2.3}
\int_{\rrb^{2}} H_m (x+\I\!y) \overline{H_n (x+\I\!y)}\; \mathrm{e}^{-(1-\alpha)x^2 - (\frac 1\alpha -1)y^2}\, \mathrm{d}x \mathrm{d}y = \frac{\pi \sqrt{\alpha}}{1-\alpha}\left( 2\, \frac {1+\alpha}{1-\alpha}\right)^n n!\,\delta_{mn},
\end{equation}
where  $0 < \alpha < 1$ is a parameter. The space $\mathcal{X}^{\alpha}_{\rm{hol,1}}$ of entire functions $f$ such that
\begin{equation*}
\int_{\rrb^2} |f(z)|^2 {\rm e}^{\alpha x^2 - \frac{1}{\alpha} y^2}  \D x\, \D y < \infty, \quad z = x + {\rm i}y, 
\end{equation*} 
is a Hilbert space with the inner product 
\begin{equation*}
\langle f, h \rangle \okr \int_{\rrb^{2}} f(z) \overline{h(z)} \E^{\alpha x^{2} - \frac{1}{\alpha}y^{2}} \D x \D y,
\end{equation*}
in which $h_{n}^{(\alpha)}(z)$ defined by 
\begin{equation*}
h^{(\alpha)}_{ n}(z) \okr {\rm e}^{-\frac{z^2}{2}} \left[\frac{\pi \sqrt{\alpha}}{1-\alpha}\left( 2\, \frac {1+\alpha}{1-\alpha}\right)^n n!\right]^{-\frac{1}{2}}H_{n}(z),\quad z\in\mathbb C, 
\end{equation*}
constitute, due to \eqref{GHSz:eq2.3}, an orthonormal basis. Moreover, because
\begin{equation*}
\sum_{n=0}^{\infty} \big\vert  h^{(\alpha)}_{n}(z)\big\vert^{2} < +\infty,
\end{equation*}
the formula \eqref{GHSz:f1.21.07} allows us to initiate Zaremba's procedure ensuring that the space $\mathcal{X}^{\alpha}_{\rm{hol,1}}$ is RKHS with the kernel
\begin{align*}
\begin{split}
K^{(\alpha)}(z, w) & \okr \sum_{n=0}^\infty h^{(\alpha)}_{n}(z)\, \overline{h^{(\alpha)}_{n}(w)} \\
&= \E^{-\frac{z^2+\sbar{w}^2}{2}}\frac {1-\alpha^2}{2\pi \alpha} \exp \left[ -\frac {(1-\alpha)^2}{4\alpha} (z^2 + \bar{w}^2) + \frac {1-\alpha^2}{2\alpha}z \bar{w}\right],\; z,w\in\mathbb C.
\end{split}
\end{align*}

\paragraph{\sc The Segal-Bargmann transform}
By the classical Bargmann space  $\mathcal{H}_{\rm{hol},1}$ \cite{GHSz:VBargmann61} we mean the space of those functions in $\mathcal{L}^{2}(\mathbb C, \pi^{-1}{\mathrm{e}^{-\vert z\vert^2}}\mathrm{d}x\,\mathrm{d}y )$ which are entire, or equivalently, those which are the closure of all polynomials $\mathbb{C}[Z]$ in $\mathcal{L}^{2}(\mathbb C, \pi^{-1}{\mathrm{e}^{-\vert z\vert^2}}\mathrm{d}x\,\mathrm{d}y )$. Recall that the monomials
\begin{equation*}
\Phi_{n} (z) \okr \frac {z^n}{\sqrt{n!}}\;,\quad z\in\mathbb C, \quad n =0,1,2, \ldots,
\end{equation*}
are an orthonormal basis in $\mathcal{H}_{\rm{hol},1}$. The unitary transform (namely the Segal-Bargmann one) between $\hhc_{{\rm hol},1}$ and the physical space $\llc^{2}(\rrb)$, in which the functions
\begin{equation*}
\psi_{n}(q) = \frac{\sqrt{a}}{\sqrt{2^{n}n!\sqrt{\pi}}} \E^{-a^{2} q^{2}/2} H_{n}(a q) 
\end{equation*}
are the orthonormal basis\,\footnote{\, The mass of one-dimenional harmonic oscillator is denoted by $M$, its frequency by $\omega$, $\hbar=1$, and $a=\sqrt{M\omega}$.}, is given as an integral transform with the kernel
\begin{equation}\label{GHSz:26/07-1}
A_{1}(q, \sbar{z}) = \frac{\sqrt{a}}{\pi^{1/4}} \E^{\sqrt{2} a q \sbar{z} - \frac{1}{2}(\sbar{z}^{2} + a^{2} q^{2})}.
\end{equation}
The similar Segal-Bargmann-like transform between $\hhc_{{\rm hol},1}$ and $\xxc_{{\rm hol},1}^{\alpha}$ was found in \cite{GHSz:STAli14a}. It is shown there that the mapping $h_{n}^{(\alpha)} \mapsto \Phi_{n}$  
\begin{equation*}
\Phi_{n}(z) = (B_{1} h^{(\alpha)}_{n}) (z) = \int_{\rrb^2} B_{1}(z, \bar{w}) h^{\alpha}_{n}(w) {\rm e}^{\alpha u^2 - \frac{1}{\alpha} v^2} {\rm d}u {\rm d} v, \quad w = u + {\rm i} v,
\end{equation*}
with 
\begin{equation}\label{GHSz:eq22/07-3}
B_{1}(z, \bar{w}) = \sum_{n=0}^{\infty} \Phi_{n}(z) \overline{h^{(\alpha)}_{n}(w)} = \left(\frac{1-\alpha}{\pi \sqrt{\alpha}}\right)^{\frac{1}{2}} {\rm e}^{\sqrt{2 \epsilon} z \bar{w} - \frac{1}{2} (\epsilon z^2 + \bar{w}^2)},\quad z,w\in\mathbb C,
\end{equation}
is unitary; the notation $\epsilon = (1-\alpha)/(1+\alpha)$ is adopted here and in what follows. 

\begin{remark}\label{GHSz:28/07-1}
In constructing the transformation between the physical space $\llc^{2}(\rrb)$ and $\xxc^{\alpha}_{{\rm hol},1}$ we compose the above mappings and end up with the kernel $C_{1}(q, \sbar{w})$ 
\begin{align*}
\begin{split}
C_{1}(q, \sbar{w}) & = \int_{\ccb} A_{1}(q, \sbar{z}) B_{1}(z, \sbar{w}) \E^{-|z|^{2}} \frac{\D z}{\pi} \\
& = \frac{\sqrt{a}}{\pi^{1/4}} \left(\!\frac{1-\alpha^{2}}{2\pi\alpha\sqrt{\alpha}}\!\right)^{\frac{1}{2}} \exp\left[-\frac{1}{2\alpha}(a^{2} q^{2} + \sbar{w}^{2}) + \frac{\sqrt{1-\alpha^{2}}}{\alpha} a q \sbar{w}\right],
\end{split}
\end{align*}
which defines the unitary mapping via the integral transformation.
\end{remark}

\medskip

\paragraph{\sc Limits}

\begin{svgraybox}
The van Eijndhoven-Meiers picture enjoys interesting limit properties: $\alpha\to 0+$ and $\alpha\to 1-$. These two passages produce very different effects and must be treated separately. Having in mind our main purpose here we shall restrict ourselves to the case $\alpha\to 1-$ only. This limit preserves the crucial property needed for our construction of CSs: the existence of a suitable RKHS. 

On the other hand the limit $\alpha\to 0+$ leads to results which forbide to construct any kind of CSs being well-defined within our scheme. This is because performing this limit breaks down the fundamental condition \eqref{GHSz:f1.21.07} and, consequently, the normalizability of  CSs \cite{GHSz:KGorska17-arXiv, GHSz:AWunsche15}. Nevertheless, the polynomials $H_{m,n}(z, {\bar z})$, which arise in the limit $\alpha\to 0+$ of the two variable generalization of the van Eijndhoven-Meiers picture (see the Section 2.2),  have found plenty of  interesting applications:  to mention investigation of their relation with the entangled (in particular EPR) states begun more than 20 years ago \cite{GHSz:FanKlauder94}, continued in \cite{GHSz:FanLu04},  and still being the subject of extensive research (cf. references in footnote \footref{GHSz:przypis1}).
\end{svgraybox}

{\sc Limit $\alpha\to 1-$} 

In order to make the limit procedure more efficient we redefine the holomorphic Hermite functions $h^{(\alpha)}_{n}(z)$ as
\begin{align}\label{GHSz:24/07-20}
\begin{split}
k^{(\alpha)}_{n}(z) & \okr \left(\!\frac{2\sqrt{\alpha}}{1+\alpha}\!\right)^{\!\frac{1}{2}}\! \left[\frac{1-\alpha}{2(1+\alpha)}\right]^{\frac{n}{2}} \frac{1}{\sqrt{n!}} \E^{\frac{1-\alpha}{1+\alpha} \frac{z^{2}}{2}} H_{n}\left(\!\!\sqrt{\ulamek{2\alpha}{1-\alpha^{2}}} z\!\right) \\
& = \sqrt{\frac{2\pi\alpha}{1-\alpha^{2}}} \E^{\frac{1+\alpha^{2}}{1-\alpha^{2}} \frac{z^{2}}{2}} h_{n}^{(\alpha)}\left(\!\!\sqrt{\ulamek{2\alpha}{1-\alpha^{2}}} z\!\right).
\end{split}
\end{align}
Their orthogonality relation 
\begin{equation*}
\int_{\ccb} k_{n}^{(\alpha)}(z) \overline{k_{m}^{(\alpha)}(z)} \E^{-|z|^{2}} \frac{\D z}{\pi} = \delta_{nm}
\end{equation*}
can be immediately derived from \eqref{GHSz:eq2.3}. The reproducing kernel coincides with Bargmann's 
\begin{equation*}
K_{1}^{(\alpha)}(z, w) \okr \sum_{n=0}^{\infty} k_{n}^{(\alpha)}(z) \overline{k_{n}^{(\alpha)}(w)} = \E^{z \sbar{w}}.
\end{equation*}
The Segal-Bargmann transform $\psi_{n}(q)\mapsto k_{n}^{(\alpha)}(z)$ now is based on the kernel 
\begin{align*}
\begin{split}
\hat{C}_{1}(q, \sbar{z}) & = \sqrt{\frac{2\pi\alpha}{1-\alpha^{2}}} \E^{\frac{1+\alpha^{2}}{1-\alpha^{2}} \frac{\sbar{z}^{2}}{2}} C_{1}\!\left(\!q, \sqrt{\ulamek{2\alpha}{1-\alpha^{2}}}\sbar{z}\!\right) \\
& = \frac{\sqrt{a}}{(\pi\alpha)^{1/4}} \exp\left[-\frac{1}{2\alpha}(a^{2} q^{2} + \alpha\sbar{z}^{2}) + \sqrt{\frac{2}{\alpha}} a q \sbar{z},\right]
\end{split}
\end{align*}
which tends to the kernel \eqref{GHSz:26/07-1} when $\alpha\to 1-$. 

\noindent The last unnumbered formula on p. 97 of \cite{GHSz:SJLvanEijndhoven90} with $t = \sqrt{(1-\alpha^{2})/2\alpha}$ yields
\begin{equation}\label{GHSz:26/07-7}
\lim_{\alpha\to 1-} k^{(\alpha)}_{n}(z) = \Phi_{n}(z),
\end{equation}
i.e. recovers the Bargmann basis. Details can be found in \cite{GHSz:analytic}.

\subsection{Holomorphic Hermite polynomials in two variables}\label{GHSz:sec:2_1}

Hermite polynomials in two complex variables are defined as 
\begin{equation}\label{GHSz:16/07-1}
H_{m, n}(z_{1}, z_{2}) \okr \sum_{k=0}^{\min\{m, n\}} \binom{m}{k} \binom{n}{k} (-1)^{k} k! z_{1}^{m-k} z_{2}^{n-k}, 
\end{equation}
where $m, n=0, 1, 2, \ldots$. The essence now is to think of them as holomorphic Hermite polynomials  in $z_{1}, z_{2}\in\ccb$\,\footnote{\;Polynomials $H_{m, n}(z, \sbar{z})$ (mentioned previously as a special case of polynomials $H_{m, n}(z_{1}, z_{2})$ for $z_{1} = z$ and $z_{2}=\sbar{z}$), have been present in mathematical and physical literature for around 65 years and bear the names: Ito's polynomials, incomplete or 2D Hermite polynomials, complex Hermite polynomials,  Laguerre polynomials in two variables and possibly the other - for recent literature on the subject see  \cite{GHSz:STAli15,GHSz:STAli14,GHSz:NCotfas10, GHSz:GDattoli97,GHSz:Fan2015,GHSz:Ghanmi1,GHSz:ismailegipt,GHSz:LvFan15}).  
Here we want to emphasize that \eqref{GHSz:16/07-1}  defines polynomials in two complex variables with real coefficients while $H_{m, n}(z, \sbar{z})$ are polynomials in two real variables $x=\Re{z}$ and $y=\Im{z}$ with complex coefficients. This may be somehow confusing when the term "complex Hermite polynomials" appears for the latter, for more discussion see the introductory section in \cite{GHSz:KGorska17-arXiv}.\label{GHSz:przypis1}}.

\noindent The polynomials $H_{m, n}(z_{1}, z_{2})$ come from the generating function
\begin{equation}\label{GHSz:16/07-2}
\exp{(z_{1} s + z_{2} t -st)}=\sum_{m, n}^{\infty} \frac{s^{m} t^{n}}{m! n!} H_{m, n}(z_{1}, z_{2}).  
\end{equation}
It does \underline{not} factorize as a product of two functions which may be generating functions of two other systems of orthogonal polynomials; the lack of factorization can be seen from the operational (raising and lowering) relations
\begin{align*}\label{GHSz:19/07-1}
\begin{split}
H_{m+1, n}(z_{1}, z_{2}) = (z_{1} - \partial_{z_{2}}) H_{m, n}(z_{1}, z_{2}), &\quad  H_{m, n+1}(z_{1}, z_{2}) = (z_{2} - \partial_{z_{1}}) H_{m, n}(z_{1}, z_{2}), \\
\partial_{z_{2}} H_{m, n}(z_{1}, z_{2}) = n H_{m, n-1}(z_{1}, z_{2}), &\quad \partial_{z_{1}} H_{m, n}(z_{1}, z_{2}) = m H_{m-1, n}(z_{1}, z_{2}).
\end{split}
\end{align*}

Using \eqref{GHSz:16/07-2} the Hermite polynomials in two variables can be expressed, like it is shown in \cite[Eq.(13)]{GHSz:KGorska17-arXiv}, in terms of the Hermite polynomials in a single variable \eqref{GHSz:eq2.1}
\begin{equation}\label{GHSz:19/07-2}
H_{m, n}(z_{1}, z_{2}) = 2^{-(m+n)} \sum_{k=0}^{m} \sum_{l=0}^{n} \I^{m-k} (-\I)^{n-l} H_{k+l}(\ulamek{z_{1} + z_{2}}{2}) H_{m+n-k-l}(\ulamek{z_{1}-z_{2}}{2\I}),
\end{equation}
which does not undermine the just mentioned lack of factorizability.

Using \eqref{GHSz:19/07-2} and formula (0.5)  in \cite{GHSz:SJLvanEijndhoven90} provides us with the orthogonality relations 
\begin{multline}\label{GHSz:19/07-3}
\int_{\ccb^{2}} H_{m, n}(z_{1}, z_{2}) \overline{H_{p, q}(z_{1}, z_{2})} \exp\left({-\frac{1-\alpha}{4} |\sbar{z}_{2} + z_{1}|^{2} - \frac{1-\alpha}{4\alpha}|\sbar{z}_{2} - z_{1}|^{2}}\right) \D\, z_{1} \D\, z_{2} \\ = \frac{\pi^{2}\alpha}{(1-\alpha)^{2}} \left(\frac{1+\alpha}{1-\alpha}\right)^{m+n} m! n! \delta_{m, p} \delta_{n, q}
\end{multline}
valid for $0 < \alpha < 1$, \cite[Eq. (19)]{GHSz:KGorska17-arXiv}. Though algebraic properties of $H_{m, n}(z_{1}, z_{2})$ have been widely considered in many papers (cf. references in \footref{GHSz:przypis1}), investigation of their analytic properties done in  \cite{GHSz:KGorska17-arXiv}  is, according to our best knowledge, a \underline {novelty}.

The orthogonality relations \eqref{GHSz:19/07-3} allow to introduce {\em normalized holomorphic Hermite functions} 
\begin{equation*}
h^{(\alpha)}_{m, n}(z_{1}, z_{2}) = \frac{ 1-\alpha}{\pi\sqrt{\alpha}} \left(\frac{1-\alpha}{1+\alpha}\right)^{\frac{m+n}{2}} \frac{\exp(-\frac{z_{1}z_{2}}{2})}{\sqrt{m! n!}}\, H_{m, n}(z_{1}, z_{2}),
\end{equation*}
where $z_{1}, z_{2}\in\ccb$ and $0 < \alpha < 1$. These functions satisfy the relation
\begin{equation*}
\sum_{m, n=0}^{\infty} \big\vert h_{m, n}^{(\alpha)}(z_{1}, z_{2})\big\vert^{2} = \frac{(1-\alpha^2)}{4\pi^{2}\alpha^{2}} \E^{-\frac{1+\alpha^{2}}{4\alpha}(z_{1} z_{2} +\sbar{z}_{1}\sbar{z}_{2}) + \frac{1-\alpha^{2}}{4\alpha}(z_{1}\sbar{z}_{1} + z_{2}\sbar{z}_{2})} < +\infty,
\end{equation*}
which according to Zaremba's procedure makes it possible to introduce $\hhc^{(\alpha)}$, being a  RKHS with the kernel 
\begin{align*}
\begin{split}
& {K}^{(\alpha)}(z_{1}, z_{2}, w_{1}, w_{2}) = \sum_{m, n=0}^{\infty} h^{(\alpha)}_{m, n}(z_{1}, z_{2}) \overline{h^{(\alpha)}_{m, n}(w_{1}, w_{2})} \\
& \quad= \frac{(1-\alpha^{2})^{2}}{4\pi^{2} \alpha^{2}} \exp{\left[\frac{1-\alpha^{2}}{4\alpha}(z_{1} \sbar{w}_{1} + z_{2} \sbar{w}_{2}) - \frac{1+\alpha^{2}}{4\alpha}(z_{1} z_{2} + \sbar{w}_{1}\sbar{w}_{2})\right]},
\end{split}
\end{align*}
calculated using either the formula (26) and Lemma 8 in \cite{GHSz:KGorska17-arXiv} or  \cite[formula (5.2)]{GHSz:AWunsche15}.

\paragraph{\sc The Segal-Bargmann transform}

The monomials 
\begin{equation*}
\Phi_{m, n}(z_{1}, z_{2}) \okr \frac{z_{1}^{m}}{\sqrt{m!}} \frac{z_{2}^{n}}{\sqrt{n!}}, \quad z_{1}, z_{2}\in\ccb, \quad m, n=0, 1, 2, \ldots,
\end{equation*}
form an orthonormal basis in the two variable Bargmann space $\hhc_{\rm hol,2} \okr {\rm Hol}(\ccb^{2})\cap \llc^{2}(\ccb^{2}, \pi^{-2} \exp{(-|z_{1}|^{2} - |z_{2}|^{2})} \D\, z_{1} \D\, z_{2})$
and may be transformed into the square integrable functions describing the system of two independent harmonic oscillators
\begin{align*}
\begin{split}
\psi_{m, n}(q_{1}, q_{2}) & = \psi_{m}(q_{1}) \psi_{n}(q_{2}) \\&= \frac{\sqrt{ab}}{\sqrt{2^{m+n} m! n! \pi}} \E^{-(a^{2} q_{1}^{2} + b^{2} q_{2}^{2})/2} H_{m}(a q_{1}) H_{n}(b q_{2}),
\end{split}
\end{align*}
which are the basis in the physical Hilbert space $\llc^{2}(\rrb^{2})$\,\footnote{\;We take that $\hbar=1$, $a = \sqrt{M \omega_{x}}$ and $b = \sqrt{M \omega_{y}}$, $M$ denotes the total mass of the system and frequencies of oscillations in $x$ and $y$ direction are $\omega_{x}$ and $\omega_{y}$, respectively.}. The mapping $\hhc_{{\rm hol}, 2}$ into $\llc^{2}(\rrb^{2})$ is unitary and has the kernel 
\begin{align}\label{GHSz:20/07-5}
\begin{split}
A_{2}(q_{1}, q_{2}, \sbar{z}_{1}, \sbar{z}_{2}) & = A_{1}(q_{1}, \sbar{z}_{1}) A_{1}(q_{2}, \sbar{z}_{2}) \\
& = \sqrt{\frac{ab}{\pi}} \E^{-\frac{1}{2}(\sbar{z}_{1}^{2} + \sbar{z}_{2}^{2}) -\frac{1}{2}(a^{2} q_{1}^{2} + b^{2} q_{2}^{2})} \E^{\sqrt{2}(a q_{1} \sbar{z}_{1} + b q_{2} \sbar{z}_{2}))}.
\end{split}
\end{align}
Thus, we have
\begin{equation*}
\Phi_{m, n}(z_{1}, z_{2}) = \int_{\rrb^{2}} \psi_{m, n}(q_{1}, q_{2}) A_{2}(q_{1}, q_{2}, \sbar{z}_{1}, \sbar{z}_{2}) \D\,q_{1} \D\,q_{2}.
\end{equation*}
Another Bargman-like transform acts between the spaces $\hhc^{(\alpha)}$ and $\hhc_{{\rm hol}, 2}$. It has been shown in \cite[{Section ``Relating $\hhc^{(\alpha)}$ to the Bargmann space''}]{GHSz:KGorska17-arXiv} that this transform is also unitary and possesses the kernel 
\begin{align*}
\begin{split}
{B}_{2}(z_{1}, z_{2}, \sbar{w}_{1}, \sbar{w}_{2}) & = \sum_{m, n=0}^{\infty} \Phi_{m, n}(z_{1}, z_{2}) \overline{h^{(\alpha)}_{m, n}(w_{1}, w_{2})} \\
&= \frac{1-\alpha}{\pi\sqrt{\alpha}} \E^{-\frac{1}{2} \sbar{w}_{1} \sbar{w}_{2} + \sqrt{\epsilon}(z_{1} \sbar{w}_{1} + z_{2} \sbar{w}_{2}) - \epsilon z_{1} z_{2}},
\end{split}
\end{align*}
where $\epsilon$ was defined just after \eqref{GHSz:eq22/07-3} and $0 < \alpha < 1$. That leads to
\begin{equation*}
h^{(\alpha)}_{m, n}(w_{1}, w_{2}) = \int_{\ccb^{2}} \Phi_{m, n}(z_{1}, z_{2}) B_{2}(z_{1}, z_{2}, \sbar{w}_{1}, \sbar{w}_{2}) \E^{-|z_{1}|^{2} - |z_{2}|^{2}} \frac{\D\,z_{1} \D\,z_{2}}{\pi^{2}}.
\end{equation*}

Extending Remark \ref{GHSz:28/07-1} 
to the 2D case we compose the transformations $A_{2}$ and $B_{2}$ and obtain the unitary mapping $\llc^{2}\mapsto \hhc^{(\alpha)}$ with the kernel 
\begin{align*}
\begin{split}
& {C}_{2}(q_{1}, q_{2}, \sbar{w}_{1}, \sbar{w}_{2}) = \int_{\ccb^{2}} A_{2}(q_{1}, q_{2}, \sbar{z}_{1}, \sbar{z}_{2}) {B}_{2}(z_{1}, z_{2}, \sbar{w}_{1}, \sbar{w}_{2}) \E^{-|z_{1}|^{2} - |z_{2}|^{2}} \frac{\D\,z_{1} \D\,z_{2}}{\pi^{2}} \\
& \quad = \sqrt{\frac{ab}{\pi}} \frac{1-\alpha^{2}}{2\pi\alpha} \E^{-\frac{1+\alpha^{2}}{4\alpha} (a^{2} q_{1}^{2} + b^{2} q_{2}^{2}) - \frac{(1-\alpha)^{2}}{2\alpha} ab q_{1} q_{2}} \E^{-\frac{1-\alpha^{2}}{8\alpha}(\sbar{w}_{1}^{2} + \sbar{w}_{2}^{2})}\\
& \quad \times  \E^{- \frac{1+\alpha^{2}}{4\alpha} \sbar{w}_{1} \sbar{w}_{2}} \E^{\frac{\sqrt{2(1-\alpha^{2})}}{4\alpha}[(1+\alpha)(aq_{1} \sbar{w}_{1} + bq_{2} \sbar{w}_{2}) + (1-\alpha)(aq_{1} \sbar{w}_{2} + bq_{2} \sbar{w}_{1})]}.
\end{split}
\end{align*}

\paragraph{\sc Limit $\alpha\to 1-$} 
The limit case $\alpha\to 1-$ will be considered analogously to what was done for the Hermite polynomials in a single variable. We begin with redefining Hermite functions in two variables $h^{(\alpha)}_{m, n}(z_{1}, z_{2})$ as follows
\begin{align*}
\begin{split}
k^{(\alpha)}_{m, n}(z_{1}, z_{2}) & \okr \frac{2\sqrt{\alpha}}{1+\alpha} \left(\frac{1-\alpha}{1+\alpha}\right)^{\!\!\frac{m+n}{2}} \frac{\exp\big(\frac{1-\alpha}{1+\alpha} z_{1} z_{2}\!\big)}{\sqrt{m! n!}} H_{m, n}\left(\!\ulamek{2\sqrt{\alpha} z_{1}}{\sqrt{1-\alpha^{2}}}, \ulamek{2\sqrt{\alpha} z_{2}}{\sqrt{1-\alpha^{2}}}\!\!\right) \\
& = \frac{2\pi\alpha}{1-\alpha^{2}} \exp{(\ulamek{1+\alpha^{2}}{1-\alpha^{2}} z_{1} z_{2})} h^{(\alpha)}_{m, n}\left(\ulamek{2\sqrt{\alpha} z_{1}}{\sqrt{1-\alpha^{2}}}, \ulamek{2\sqrt{\alpha} z_{2}}{\sqrt{1-\alpha^{2}}}\right).
\end{split}
\end{align*}

They satisfy the orthogonality relation
\begin{equation*}
\int_{\ccb^{2}} k^{(\alpha)}_{m, n}(z_{1}, z_{2}) \overline{k^{(\alpha)}_{m, n}(z_{1}, z_{2})} \exp{(-|z_{1}|^{2} - |z_{2}|^{2})} \frac{\D\, z_{1} \D\, z_{2}}{\pi^{2}} = \delta_{m, p} \delta_{n, q}
\end{equation*}
and form RKHS $\kkc^{(\alpha)}$ with the kernel
\begin{equation*}
K^{(\alpha)}(z_{1}, z_{2}, w_{1}, w_{2}) = \exp(z_{1} \sbar{w}_{1} + z_{2} \sbar{w}_{2}),
\end{equation*}
which again coincides with the two dimensional Bargmann one. 
The Segal-Bargmann transform connecting the spaces $\llc^{2}(\rrb^{2})$ and $\kkc^{(\alpha)}$ reads
\begin{align}
\begin{split}
\hat{C}_{2}(x, y, \sbar{w}_{1}, \sbar{w}_{2}) & = \frac{2\pi\alpha}{1-\alpha^{2}} \E^{\frac{1+\alpha^{2}}{1-\alpha^{2}}\sbar{w}_{1} \sbar{w}_{2}} C_{2}\left(q_{1}, q_{2}, \ulamek{2\sqrt{\alpha}}{\sqrt{1-\alpha^{2}}} \sbar{w}_{1}, \ulamek{2\sqrt{\alpha}}{\sqrt{1-\alpha^{2}}} \sbar{w}_{2}\right) \\
&= 
\sqrt{\frac{a b}{\pi}} \E^{-\ulamek{1+\alpha^{2}}{4\alpha}(a^{2} q_{1}^{2} + b^{2} q_{2}^{2})} \E^{-\ulamek{1}{2}(\sbar{w}_{1}^{2} + \sbar{w}_{2}^{2})} \E^{-\ulamek{1-\alpha^{2}}{2\alpha} a b q_{1} q_{2}} \\
& \times \E^{\ulamek{1+\alpha}{\sqrt{2\alpha}}(aq_{1} \sbar{w}_{1} + bq_{2} \sbar{w}_{2})} \E^{\ulamek{1-\alpha}{\sqrt{2\alpha}}(aq_{1} \sbar{w}_{2} + b q_{2}\sbar{w}_{1})},
\end{split}\notag
\end{align}
and in the limit $\alpha\to 1-$ tends to \eqref{GHSz:20/07-5}. Analogously to \eqref{GHSz:26/07-7} one gets
\begin{equation*}\label{GHSz:31/07-1}
\lim_{\alpha\to 1-} k^{(\alpha)}_{m,n}(z_1,{z_2}) = \Phi_{m,n}(z_1,z_2),
\end{equation*}
i.e. performing the limit procedure we end up on the 2D Bargmann basis.

\section{HSz CSs - holomorphic Hermite polynomials perspective}\label{GHSz:sec.30}

Let us recall that, according to our definition of coherent states evolved in Section 1.4 and starting from the formula \eqref{GHSz:23/07-2}, the basic requirement for some states to be called coherent is to be provided with $(\Phi_n)_{n} $ satisfying \eqref{GHSz:f1.21.07}. Now the Zaremba construction guarantees existence of the Segal-Bargmann transform, the property which is historically and not too rigorously identified with the overcompleteness and/or the resolution of the identity.

\begin{svgraybox}
\noindent Our definition of coherent states allows to put traditional (like those in A,B and C on page \pageref{AAA}) coherent and squeezed states on the same footing.
\end{svgraybox} 

\noindent Considerations in the previous section show that introduced there complex Hermite functions fulfill the conditions imposed on  RKHS. This opens a green light to engage it in construction of the coherent states.  We shall do it in the next Section and show that so obtained states \underline {not only satisfy the resolution of the identity} (which by the way is resulting from their construction) but they turn out to be also entangled.  To make the above statement more precise let us formulate the definition of what has to be understand under the notion of entanglement.
\begin{svgraybox}
Suppose two separable Hilbert spaces $\hhc$ and $\kkc$ are given. Let $\hhc\bigotimes\kkc$ be the state space. Call a state $c$ in $\hhc\bigotimes\kkc$ {\em decomposable} (or {\em factorizable}) if $c=c_{\hhc}\otimes c_{\kkc}$ with $c_{\hhc}\in\hhc$ and $c_{\kkc}\in\kkc$. A state which is not decomposable will be called {\em entangled}. Referring to CSs we can say that the family $\{c_{x}\}_{x\in X}$ is decomposable if 
\begin{gather}
c_{x}=\sum_{m,n=0}^{\infty}\Phi_{m}(x)e_{m}\otimes\Psi_{n}(x)f_{n},\;x\in X,\notag
\\
\text{$(\Phi_{m})_{m=0}^{\infty}$ and $(\Psi_{n})_{n=0}^{\infty}$ are orthonormal bases in suitable RKHSs.}\notag
\end{gather}

%
\begin{minipage}{300pt}
If there are \underbar{no} such $(\Phi_{m})_{m=0}^{\infty}$ and $(\Psi_{n})_{n=0}^{\infty}$ making the above decomposition possible the family becomes entangled by definition\footnote{In the literature there is no undoubtedly defined notion of the entanglement (cf. footnote \footref{bell}).}. However,  $\{c_{x}\}_{x\in X}$ as members of the state space $\hhc\bigotimes\kkc$ are HSz CSs \underbar{anyway}.
\end{minipage}
\end{svgraybox}

In what follows we will provide the reader with a keystone  example of new bosonic states, based on the holomorphic Hermite polynomials, which are coherent and entangled simultaneously\,\footnote{\;\label{bell}States which are frequently appearing in the literature under the name {\em coherent entangled states} are bipartite Bell-like states constructed using  tensor products of standard coherent states, usually  $|z\rangle$ and $|-z\rangle$ \cite{GHSz:Sanders}; they are obviously entangled but not coherent in any commonly acceptable  sense.}. Even more, these new states appear to be squeezed. Surprisingly, the limit procedures which the Hermite polynomials enjoy {\bf allow to link entangled and decomposable states within the HSz coherent states framework}.  All this happens, due to the Proposition \ref{propro}, under the guidance of HSz coherent states merging mathematical and physical aspects of the novel CSs. \label{?}

\section{CSs for holomorphic Hermite polynomials.}\label{GHSz:sec:3}
\subsection{Single particle Hermite CSs - coherence and squeezing}

Single particle CSs corresponding to the sequence $(k_{n}^{(\alpha)})_{n}$ are defined as
\begin{equation*}
c_{z}^{(\alpha)} = \sum_{n=0}^{\infty} k_{n}^{(\alpha)}(z) e_{n}, \quad z\in\ccb.
\end{equation*}
Using the recurrence relation $H_{n+1} = 2z H_{n}(z) - 2nH_{n-1}(z)$ one gets 
\begin{equation*}
k_{n+1}^{(\alpha)}(z) = z \frac{2\sqrt{\alpha}}{1+\alpha} \frac{1}{\sqrt{n+1}} k_{n}^{(\alpha)}(z) - \frac{1-\alpha}{1+\alpha} \sqrt{\frac{n}{n+1}} k_{n-1}^{(\alpha)}(z)
\end{equation*}
and shows that $c_{z}^{(\alpha)}$ appear to be eigenfunctions
\begin{equation*}
B_{-}c_{z}^{(\alpha)}= zc_{z}^{(\alpha)}
\end{equation*}
 of the operator 
\begin{equation}\label{GHSz:26/07-10}
B_{-} \okr 
\frac{1+\alpha}{2\sqrt{\alpha}} b + \frac{1-\alpha}{2\sqrt{\alpha}} b^{\dag}, 
\end{equation}
where $b$ and $b^{\dag}$ denote the canonical annihilation and creation operators. $B_{-}$ together with $B_{+}$ given by 
\begin{equation}\label{GHSz:26/07-11}
B_{+} \okr  
\frac{1+\alpha}{2\sqrt{\alpha}} b^{\dag} + \frac{1-\alpha}{2\sqrt{\alpha}} b
\end{equation}  
satisfy the commutation relations $[B_{-}, B_{+}] = 1$ and $[B_{-}, B_{-}] = [B_{+}, B_{+}] =0$ which mean that \eqref{GHSz:26/07-10} and \eqref{GHSz:26/07-11} belong to the class of the Bogolubov transformations \cite{GHSz:FetterWalecka}, the relation of which to the squeezed (coherent)  states is well established \cite{GHSz:Bogoliubov}. So the states $c_{z}^{(\alpha)}$, primarily required only to satisfy the resolution of the identity, are also squeezed states in the sense of (A), p. \pageref{GHSz:fA}, and in the limit $\alpha\to 1-$ become exclusively coherent in the traditional meaning.

\paragraph{\sc Hermite CSs and single mode squeezing operation}

The squeezed states $\eta_{z}^{\xi}$ may also be introduced through {\em squeezing operation} acting on the standard coherent states 
\begin{equation*}
\eta_{z}^{\xi} \okr S(\xi) \eta_{z}, \quad S(\xi) = \E^{\xi K_{+} - \sbar{\xi} K_{-}}, \quad \xi\in\ccb,
\end{equation*}
where $K_{\pm}$ are the generators of $su(1, 1)$  algebra which, together with the third one $K_{0}$, satisfy the commutation relations
\begin{equation*}
[K_{-}, K_{+}] = 2K_{0}, \quad [K_{0}, K_{\pm}] = \pm K_{\pm}.
\end{equation*}
Setting $\zeta = \xi \tanh(|\xi |)/|\xi |$, $|\zeta| < 1$, the squeeze operator $S(\xi)$ can be disentangled employing the well-known Zassenhaus formula
\begin{equation}\label{GHSz:24/07-8}
S(\xi) = \E^{\xi K_{+} - \sbar{\xi} K_{-}} = \E^{\zeta K_{+}} \E^{\ln(1+|\zeta|^{2}) K_{0}} \E^{-\sbar{\zeta} K_{-}}.
\end{equation}
This may be used to obtain so-called {\em squeezed basis} $e_{n}^{\xi} \okr S(\xi) e_{n}$, $n=0, 1, 2, \ldots$ with which
the squeezed states $\eta_{z}^{\xi}$ are written as
\begin{equation*}
\eta_{z}^{\xi} = \sum_{n=0}^{\infty} \frac{{z}^{\,n}}{\sqrt{n!}} e_{n}^{\xi}. 
\end{equation*}

\noindent Since the squeeze operator is unitary the squeezed basis is also orthonormal in the Bargmann space $\hhc_{\rm hol,1}$ and squeezed states satisfy the same resolution of identity as $\eta_{z}$. 

In the Bargmann representation the operators $K_{\pm}$ and $K_{0}$ have the form
\begin{equation}\label{GHSz:24/07-10}
K_{+} = \ulamek{1}{2} z^{2}, \quad K_{-} = \ulamek{1}{2} \partial^{2}_{z}, \quad K_{0} = \ulamek{1}{2}\big(\ulamek{1}{2} + z\partial_{z}\big).
\end{equation}
The squeezed RKHS basis $\Phi_{n}^{\xi}$ is determined by the action of $S(\xi)$ (given by \eqref{GHSz:24/07-8} with \eqref{GHSz:24/07-10} put in) on $\Phi_{n}(z)$. The calculation
presented in \cite{GHSz:STAli14a} leads to 
\begin{equation}\label{GHSz:24/07-11}
\Phi_{n}^{\xi}(z) = (1-|\zeta|^{2})^{\frac{1}{4}} \E^{\frac{\zeta}{2} z^{2}} \frac{\sbar{\zeta}^{\frac{n}{2}}}{\sqrt{2^{n} n!}} H_{n}\left(\!\!\sqrt{\ulamek{1-|\zeta|^{2}}{2\sbar{\zeta}}} z\!\right).
\end{equation}
From the algebraic relation $H_{n+1} = 2 z H_{n} - H'_{n}$ we get
\begin{equation*}
\sqrt{n+1} \Phi_{n+1}^{\xi}(z) = A_{+} \Phi_{n}^{\xi}(z) \quad \text{with} \quad A_{+} = (1-|\zeta|^{2})^{-\frac{1}{2}}(z - \sbar{\zeta} \partial_{z}),
\end{equation*}
while the twin relation $2n H_{n-1} = H_{n}'$ implies
\begin{equation*}
\sqrt{n} \Phi_{n-1}^{\xi}(z) = A_{-} \Phi_{n}^{\xi}(z)\quad \text{where} \quad A_{-} = (1-|\zeta|^{2})^{-\frac{1}{2}} (\partial_{z} - \zeta z).
\end{equation*}
Assuming that $\zeta = \epsilon$ (defined below \eqref{GHSz:eq22/07-3}) we obtain, because of \eqref{GHSz:24/07-11}, that $\Phi_{n}^{\arctan(\epsilon)}(z) = k_{n}^{(\alpha)}(z)$ 
given by \eqref{GHSz:24/07-20}. That provides us the physical interpretation of the up-to-now mathematically contemplated parameter $\alpha$ \cite{GHSz:SJLvanEijndhoven90,GHSz:analytic} - from now it is to be identified with the physical squeezing parameter which measures the ratio between coordinate and momentum uncertainties.

Comparing operators $B_{-}$ and $B_{+}$ with $A_{-}$ and $A_{+}$ for $\zeta=\epsilon$ we get the Segal-Bargmann representation of operators $b$ and $b^{\dag}$ \cite{GHSz:analytic}
\begin{equation*}
b = \frac{1+\alpha^{2}}{2\alpha} \partial_{z} - \frac{1-\alpha^{2}}{2\alpha} z \quad \text{and} \quad b^{\dag} = \frac{1+\alpha^{2}}{2\alpha}z - \frac{1-\alpha^{2}}{2\alpha}  \partial_{z}.
\end{equation*}
As it should be it goes  to the standard Bargmann representation for $\alpha\to 1-$.

\subsection{Bipartite CSs - coherence, squeezing and entanglement}

\paragraph{\sc Coherent states $c_{z_{1}, z_{2}}^{(\alpha)}$}

Our approach to CSs, based on the definition given in Section 1.4, is by no means restricted to the single particle case. It may be automatically extended to multipartite systems. Here we shall present an application to bipartite systems taking as a starting point holomorphic Hermite functions in two variables $k^{(\alpha)}_{m, n}$ enabling to construct CSs. Taking as a state space $\hhc\bigotimes\kkc$, where each of $\hhc$ and $\kkc$ is a state space for itself, according to our scheme we can introduce the family of CSs
\begin{equation}\label{GHSz:22/07-4}
c_{z_{1}, z_{2}}^{(\alpha)} = \sum_{m, n} k^{(\alpha)}_{m, n}(z_{1}, z_{2}) (e_{m}\otimes f_{n}), \quad z_1,z_2\in \ccb^2,
\end{equation}
which reside in the Hilbert space $\hhc\bigotimes\kkc$.

\noindent The recurrence relations \cite[(12)]{GHSz:KGorska17-arXiv}
\begin{align*}
H_{m+1, n}(z_{1}, z_{2}) & = z_{1} H_{m, n}(z_{1}, z_{2}) - n H_{m, n-1}(z_{1}, z_{2}), \\
H_{m, n+1}(z_{1}, z_{2}) & = z_{2} H_{m, n}(z_{1}, z_{2}) - m H_{m-1, n}(z_{1}, z_{2})
\end{align*}
lead to
\begin{align}\label{GHSz:1/08-5}
\begin{split}
k_{m+1, n}^{(\alpha)}(z_{1}, z_{2}) & = z_{1} \frac{2\sqrt{\alpha}}{\sqrt{1-\alpha^{2}}} \frac{1}{\sqrt{m+1}} k_{m, n}^{(\alpha)}(z_{1}, z_{2}) - \frac{1-\alpha}{1+\alpha} \sqrt{\frac{n}{m+1}} k_{m, n-1}^{(\alpha)}(z_{1}, z_{2}), \\
k_{m, n+1}^{(\alpha)}(z_{1}, z_{2}) & = z_{2} \frac{2\sqrt{\alpha}}{\sqrt{1-\alpha^{2}}} \frac{1}{\sqrt{n+1}} k_{m, n}^{(\alpha)}(z_{1}, z_{2}) - \frac{1-\alpha}{1+\alpha} \sqrt{\frac{m}{n+1}} k_{m-1, n}^{(\alpha)}(z_{1}, z_{2}),
\end{split}
\end{align} 
which enable one to show that the states $c_{z_{1}, z_{2}}^{(\alpha)}$  are common eigenvectors 
\begin{equation}\label{GHSz:01/08-3}
B_{1, -} c_{z_{1}, z_{2}}^{(\alpha)}= z_1 c_{z_{1}, z_{2}}^{(\alpha)}\quad\quad B_{2, -}c_{z_{1}, z_{2}}^{(\alpha)}= z_2 c_{z_{1}, z_{2}},^{(\alpha)},\quad z_1,z_2\in \ccb^2,
\end{equation}
of the operators $B_{1, -}$ and $B_{2, -}$ 
\begin{equation}\label{GHSz:22/07-6}
B_{1, -} \okr \frac{1+\alpha}{2\sqrt{\alpha}} b_{1} + \frac{1-\alpha}{2\sqrt{\alpha}}b_{2}^{\dag}, \quad B_{2, -} \okr \frac{1-\alpha}{2\sqrt{\alpha}} b_{1}^{\dag} + \frac{1+\alpha}{2\sqrt{\alpha}} b_{2}
\end{equation}
where $b_{i}^{\dag}$ and $b_{i}$ ($i=1, 2$) denote the canonical creation and anihillation operators for the modes $i=1, 2$. Operators $B_{i, -}$ together with their adjoints $B_{i, +}$, $i=1, 2$  
satisfy the standard canonical commutation relations $[B_{i, -}, B_{j, +}] = \delta_{i j}$, $[B_{i, -}, B_{j, -}] = [B_{i, +}, B_{j, +}]=0$ for $i, j=1, 2$. Proceeding further and using \eqref{GHSz:01/08-3} one shows that 
\begin{equation}
\label{GHSz:01/08-8}
B_{1, -} \otimes B_{2, -} c_{z_{1}, z_{2}}^{(\alpha)} = z_1 z_2 c_{z_{1}, z_{2}}^{(\alpha)} \quad z_1,z_2\in \ccb^2.
\end{equation}
Taken together \eqref{GHSz:01/08-3} and \eqref{GHSz:01/08-8} mean that $c_{z_{1}, z_{2}}^{(\alpha)}$ fulfill the postulate (A) listed on the p. \pageref{GHSz:fA}, generalized here to the multimode case, i.e. to the set of mutually commuting operators playing the role of annihilators. Simultaneously, because of \eqref{GHSz:22/07-6}, we see that this time we deal with the Bogolubov transformation which (unlike  for the single particle case) mixes the modes. But, like previously, appearance of the Bogolubov transformation suggests that $c_{z_{1}, z_{2}}^{(\alpha)}$ may have something in common with squeezed states - this will be clarified in the next Section.

\paragraph{\sc Hermite CSs and two mode squeezing operation}

Consider the two mode representation of the generators of $su(1, 1)$ algebra given by
\begin{equation}\label{GHSz:24/07-2}
K_{+} = z_{1} z_{2}, \quad K_{-} = \partial _{z_{1}} \partial_{ z_{2}}, \quad K_{0} = \ulamek{1}{2}(1 + z_{1}\partial_{z_{1}} + z_{2} \partial_{z_{2}}),
\end{equation}
and extend the definition of the RKHS squeezed basis  to the bipartite system
\begin{equation}\label{GHSz:24/07-3}
\Phi_{m, n}^{\xi}(z_{1}, z_{2}) = S(\xi) \Phi_{m, n}(z_{1}, z_{2}), \quad \text{where} \quad S(\xi) = \E^{\xi K_{+} - \sbar{\xi} K_{-}}.
\end{equation}
Then, using  \eqref{GHSz:24/07-3}, \eqref{GHSz:24/07-8}, \eqref{GHSz:24/07-2} and \cite[(I.5.2d) on p. 24]{GHSz:GDattoli97} we get
\begin{equation*}
\Phi_{m, n}^{\xi}(z_{1}, z_{2}) = \sqrt{1-|\zeta|^{2}} \frac{\sbar{\zeta}^{\frac{m+n}{2}}}{\sqrt{m! n!}} \E^{\zeta z_{1} z_{2}} H_{m, n}\left(\sqrt{\ulamek{1-|\zeta|^{2}}{\sbar{\zeta}}} z_{1}, \sqrt{\ulamek{1-|\zeta|^{2}}{\sbar{\zeta}}} z_{2}\right),
\end{equation*}
which span the appropriate RKHS being a subspace of $\llc^{2}(\ccb^{2}, \pi^{-2}\E^{-|z_{1}|^{2} - |z_{2}|^{2}} \D z_{1} \D z_{2})$. In the Bargmann representation  creation and anihillation operators acting on the functions $f\in \lin\zb{\Phi_{m, n}^{\xi}}{m,n=0,1,\dots}$  
behave as 
\begin{align*}
\begin{split}
(A^{\zeta}_{1,+}f)(z_{1}, z_{2}) = \frac{z_{1} - \sbar{\zeta}\partial_{z_{2}}}{\sqrt{1-|\zeta|^{2}}} f(z_{1}, z_{2}), & \quad (A^{\zeta}_{1, -}f)(z_{1}, z_{2}) = \frac{\partial_{z_{1}} - \zeta z_{2}}{\sqrt{1-|\zeta|^{2}}} f(z_{1}, z_{2}), \\
(A^{\zeta}_{2, +}f)(z_{1}, z_{2}) = \frac{z_{2} - \sbar{\zeta}\partial_{z_{1}}}{\sqrt{1-|\zeta|^{2}}} f(z_{1}, z_{2}), & \quad (A^{\zeta}_{2, -}f)(z_{1}, z_{2}) = \frac{\partial_{z_{2}} - \zeta z_{1}}{\sqrt{1-|\zeta|^{2}}} f(z_{1}, z_{2})
\end{split}
\end{align*}
for $z_{1}, z_{2}\in\ccb$. 
\begin{remark}\label{GHSz:remarks3}The justification of name annihilation and creation operators comes from the fact that operators $A_{1, +/-}$ act on the first mode $m$  as
\begin{equation*}
A_{1, +}^{\zeta} \Phi_{m, n}^{\xi} = \sqrt{m+1} \Phi_{m+1, n}^{\xi}, \quad A_{1, -}^{\zeta} \Phi_{m, n}^{\xi} = \sqrt{m} \Phi_{m-1, n}^{\xi},
\end{equation*}
while $A_{2, +/-}$ act on the second mode $n$ as
\begin{equation*}
A_{2, +}^{\zeta} \Phi_{m, n}^{\xi} = \sqrt{n+1} \Phi_{m, n+1}^{\xi}, \quad A_{2, -}^{\zeta} \Phi_{m, n}^{\xi} = \sqrt{n} \Phi_{m, n-1}^{\xi}.
\end{equation*} 
\end{remark}
For $\zeta = \epsilon$ we have $\Phi_{m, n}^{\arctan(\epsilon)}(z_{1}, z_{2}) = k_{m, n}^{(\alpha)}(z_{1}, z_{2})$.
 Comparing $A^{\epsilon}_{i, +/-}$ with $B_{i, +/-}$ we find the Bargmann representation of operators $b_{i}^{\dag}$ and $b_{i}$, $i=1, 2$\begin{align*}\label{GHSz:30/07-1}
\begin{split}
b_{1}^{\dag} = \frac{1+\alpha^{2}}{2\alpha} z_{1} - \frac{1-\alpha^{2}}{2\alpha}\partial_{z_{2}}, & \quad b_{1} = \frac{1+\alpha^{2}}{2\alpha}\partial_{z_{1}} - \frac{1-\alpha^{2}}{2\alpha} z_{2}, \\
b_{2}^{\dag} = \frac{1+\alpha^{2}}{2\alpha} z_{2} - \frac{1-\alpha^{2}}{2\alpha}\partial_{z_{1}}, & \quad b_{2} = \frac{1+\alpha^{2}}{2\alpha}\partial_{z_{2}} - \frac{1-\alpha^{2}}{2\alpha} z_{1}.
\end{split}
\end{align*}
\begin{svgraybox}
We see that the parameter $\alpha$ is responsible not only for squeezing but also for mixing the modes,  one should also notice that both these effects disappear in the limit $\alpha\to 1-$.
\end{svgraybox}

\paragraph{\sc Entangled squeezed coherent states}

As said in the Section 3 the proper definition of the entanglement qualifies a state to be entangled if it is not factorizable. Because of \eqref{GHSz:19/07-2} and operational rules satisfied by polynomials $H_{m,n}(z_1,z_2)$ this is the case for the states  $c_{z_{1}, z_{2}}^{(\alpha)}$ which can not be represented as a product of factors depending separately on $z_1$ and $z_2$. But, as it has been demonstrated, $c_{z_{1}, z_{2}}^{(\alpha)}$ are simultaneously coherent/squeezed which phenomenon at first glance may seem to be a little unexpected, nevertheless is shown to be a fact possible due to the generalization of coherence presented in our study. 

Search for quantum states which are simultaneously coherent and entangled, or, more precisely, which satisfy some criteria allowing to call them coherent and entangled, is not new. Example of  such states, called coherent-entangled, was provided in \cite{GHSz:FanLu04} where the authors found explicit form of bipartite states being common eigenvectors of the center of mass coordinate operator and the difference of canonical annihilators $a_1-a_2$ and next linked superposition of these states to the standard example  illustrating entanglement, namely to the EPR states,  i.e. common eigenstates of the center of mass coordinate and relative momentum operators. Fan-Lu states, as may be seen from Eq.8 in \cite{GHSz:FanLu04}, are nonfactorizable and satisfy the formal resolution of unity (e) but are not of a finite norm which means that they break one of requirements on which our construction is based. The problem becomes analogous to that which we have roughly mentioned in the Section 2.1 when have remarked on the limit case $\alpha\to 0{+}$.  The latter problem needs a very special  and careful analysis which goes beyond the current research and this is why we have decided to exclude it from our considerations and restrict ourselves to the statement as  follows:  
\begin{svgraybox} 
As long as $0 < \alpha < 1$  the states $c^{\alpha}_{z_1,z_2}$ given by \eqref{GHSz:22/07-4} exhibit the coherence/squeezing and entanglement peacefully coexisting and, moreover, somewhat interrelated. This is possible due to the HSz approach which proposes to look at the properties of coherent states through the reproducing kernel property and which enables us to see the resolution of the identity in much wider context, especially avoiding the restrictive assumption of rotational invariance of the measure in question. Linearity which is sitting in the heart of quantum physics and which enforces us to treat all linear combinations of elementary solutions on the same footing  supports this kind of approach. A significant feature is that the limit $\alpha\to 1-$ switches off both entanglement and squeezing but does not loose anything of coherence. 
\end{svgraybox}

\begin{acknowledgement}
The work of the third author is supported by the grant of NCN (National Science Center, Poland), decision No. DEC-2013/11/B/ST1/03613.
\end{acknowledgement}

\end{document}